\documentclass[prd,preprint,superscriptaddress,showpacs,byrevtex]{revtex4}
\usepackage{bm}
\def\fsl#1{\setbox0=\hbox{$#1$}           
   \dimen0=\wd0                                 
   \setbox1=\hbox{/} \dimen1=\wd1               
   \ifdim\dimen0>\dimen1                        
      \rlap{\hbox to \dimen0{\hfil/\hfil}}      
      #1                                        
   \else                                        
      \rlap{\hbox to \dimen1{\hfil$#1$\hfil}}   
      /                                         
   \fi}                                         %
\newcommand{\be}{\begin{equation}}
\newcommand{\ee}{\end{equation}}
\newcommand{\bea}{\begin{eqnarray}}
\newcommand{\eea}{\end{eqnarray}}
\newcommand{\beq}{\begin{equation}}
\newcommand{\eeq}{\end{equation}}
\newcommand{\beqs}{\begin{eqnarray}}
\newcommand{\eeqs}{\end{eqnarray}}

\newcommand{\aslash}{A\hspace{-0.067in}\slash}

\begin{document}
\title{ Lattice QCD Method To Study Proton Spin Crisis }
\author{Gouranga C Nayak }\thanks{E-Mail: nayakg138@gmail.com}
%
%
\date{\today}
\begin{abstract}
The proton spin crisis remains an unsolved problem in particle physics. The spin and angular momentum of the partons inside the proton are non-perturbative quantities in QCD which cannot be calculated by using the perturbative QCD (pQCD). In this paper we present the lattice QCD formulation to study the proton spin crisis. We derive the non-perturbative formula of the spin and angular momentum of the partons inside the proton from the first principle in QCD which can be calculated by using the lattice QCD method.
\end{abstract}
\pacs{ 12.38.-t, 11.30.-j, 14.20.Dh, 12.38.Gc }
\maketitle
\pagestyle{plain}

\pagenumbering{arabic}

\section{ Introduction }

The spin of the proton is $\frac{1}{2}$. Since the proton consists of quarks and gluons it was expected that the spin of the quarks and gluons inside the proton add up to  give spin $\frac{1}{2}$ of the proton. The proton in motion high energy consists of quarks, antiquarks and gluons because sea quark-antiquark pairs and gluons are produced from the QCD vacuum at high energy.

For the proton in motion along z-axis the spin/helicity $\frac{1}{2}$ was expected to be
\bea
\frac{1}{2} = <p,s|{\hat S}^z_q|p,s>+<p,s|{\hat S}^z_g|p,s>
\label{hfu}
\eea
where $<p,s|{\hat S}^z_q|p,s>$ is the spin of the quarks plus antiquarks inside the proton and $<p,s|{\hat S}^z_g|p,s>$ is the spin of the gluons inside the proton. In eq. (\ref{hfu}) the $|p,s>$ is the energy-momentum eigenstate of the longitudinally polarized proton of momentum $p^\mu$ and spin $s$ and ${\hat S}^z_q$ (${\hat S}^z_g$) is the z-component of the spin vector operator ${\vec {\hat S}}_q$ (${\vec {\hat S}}_g$) of the quarks plus antiquarks (gluons) inside the proton.

However, the experimental data suggested otherwise. In 1988-89 the EMC collaboration \cite{emu} found that only a negligible fraction of the proton spin is carried by the spin of the quarks and antiquarks inside the proton which was also confirmed by the other experiments \cite{ohu}. The addition of the spin of the gluons \cite{hcu} inside the proton is not sufficient to explain the spin $\frac{1}{2}$ of the proton. At present the world data suggests that about fifty percent of the spin of the proton is due to the spin of the quarks plus antiquarks plus gluons inside the proton \cite{wdu}. Since the remaining fifty percent spin of the proton is still missing it is known as the proton spin crisis.

In order to solve this proton spin crisis it is suggested in the literature that the orbital angular momenta of the quarks, antiquarks and gluons inside the proton should be added. In the parton model it is suggested that \cite{jfu}
\bea
\frac{1}{2} = <p,s|{\hat S}^z_q|p,s>+<p,s|{\hat S}^z_g|p,s>+<p,s|{\tilde {\hat L}}^z_q|p,s>+<p,s|{\tilde {\hat L}}^z_g|p,s>
\label{jfu}
\eea
where $<p,s|{\tilde {\hat L}}^z_q|p,s>$ and $<p,s|{\tilde {\hat L}}^z_g|p,s>$ are the orbital angular momenta of the quarks plus antiquarks and gluons respectively in the light-cone gauge inside the proton.

In terms of the gauge invariant spin/angular momentum in QCD it is suggested that \cite{xfu}
\bea
\frac{1}{2} = <p,s|{\hat S}^z_q|p,s>+<p,s|{ {\hat L}}^z_q|p,s>+<p,s|{\hat J}^z_g|p,s>
\label{xjfu}
\eea
where $<p,s|{ {\hat L}}^z_q|p,s>$ is the gauge invariant orbital angular momentum of the quarks plus antiquarks inside the proton and $<p,s|{\hat J}^z_g|p,s>$ is the gauge invariant total angular momentum of the gluons inside the proton.

The eq. (\ref{xjfu}) is not correct because of the non-zero boundary surface term in QCD in the conservation equation of the angular momentum using the gauge invariant Noether's theorem in QCD  \cite{nkymu} due to the confinement of quarks and gluons inside the finite size proton \cite{nkbsu}. Because of this the angular momentum sum rule as given by eq. (\ref{xjfu}) is violated in QCD \cite{nkagu}.

By taking the confinement effect into account one finds in QCD that \cite{nkagu}
\bea
\frac{1}{2} = <p,s|{\hat S^z}_q|p,s>+<p,s|{ {\hat L}}^z_q|p,s>+<p,s|{\hat J}^z_g|p,s>+<p,s|{\hat J}^z_B|p,s>
\label{nkfu}
\eea
where $<p,s|{\hat J}^z_B|p,s>$ is the boundary surface term contribution to the total angular momentum in QCD due to the confinement of quarks and gluons inside the finite size proton \cite{nkbsu}.

Note that $<p,s|{\hat S}_q|p,s>$, $<p,s|{ {\hat L}}_q|p,s>$, $<p,s|{\hat J}_g|p,s>$ and $<p,s|{\hat J}_B|p,s>$ in eq. (\ref{nkfu}) are non-perturbative quantities which cannot be calculated by using the perturbative QCD (pQCD) method. Hence the non-perturbative QCD is necessary to calculate the quantities $<p,s|{\hat S}_q|p,s>$, $<p,s|{ {\hat L}}_q|p,s>$, $<p,s|{\hat J}_g|p,s>$ and $<p,s|{\hat J}_B|p,s>$ in eq. (\ref{nkfu}). However, the analytical solution of the non-perturbative QCD is not known yet. Hence the first principle method to study non-perturbative QCD is by using the lattice QCD method.

Recently we have presented the formulation of the lattice QCD method to study the hadron formation from quarks and gluons by incorporating the non-zero boundary surface term in QCD due to the confinement of quarks and gluons inside the finite size proton \cite{nklqu}.

In this paper we extend this to study the proton spin crisis and present the lattice QCD formulation to study the proton spin crisis by incorporating this non-zero boundary surface term in QCD due to confinement. We derive the non-perturbative formula of the $<p,s|{\hat S}_q|p,s>$, $<p,s|{ {\hat L}}_q|p,s>$, $<p,s|{\hat J}_g|p,s>$ and $<p,s|{\hat J}_B|p,s>$ in eq. (\ref{nkfu}) from the first principle in QCD which can be calculated by using the lattice QCD method by incorporating this non-zero boundary surface term due to confinement.

The paper is organized as follows. In section II we discuss the effect of confinement on angular momentum sum rule violation in QCD. In section III we discuss the proton formation from quarks and gluons by using the lattice QCD method. In section IV we present the lattice QCD formulation to study the proton spin crisis and derive the non-perturbative formula of the spin and angular momentum of the partons inside the proton from the first principle in QCD which can be calculated by using the lattice QCD method. Section V contains conclusions.

\section{ Effect of Confinement on Angular momentum sum rule violation in QCD }

The conservation equation of the angular momentum in QCD can be obtained from the first principle by using the gauge invariant Noether's theorem in QCD under the combined gauge transformation plus rotation. The continuity equation obtained from the gauge invariant Noether's theorem in QCD is given by \cite{nkymu}
\bea
\partial_\nu J^{\nu \delta \sigma}(x) =0
\label{jmu}
\eea
where $J^{\nu \delta \sigma}(x)$ is the gauge invariant angular momentum tensor density in QCD which is related to the gauge invariant and symmetric energy-momentum tensor density $T^{\nu \delta}(x)$ in QCD via the equation
\bea
J^{\nu \delta \sigma}(x) = T^{\nu \sigma}(x) x^\delta -T^{\nu \delta}(x) x^\sigma.
\label{jmnu}
\eea
The gauge invariant and symmetric energy-momentum tensor density $T^{\nu \delta}(x)$ in QCD is given by
\bea
&& T^{\nu \delta}(x) = F^{\nu \sigma b}(x)F_\sigma^{~ \delta b}(x)+\frac{1}{4} g^{\nu \delta} F^{\sigma \mu b}(x) F_{\sigma \mu}^b(x) +\frac{i}{4} {\bar \psi}_l(x)[\gamma^\nu (\delta^{lk} {\overrightarrow \partial}^\delta-igT^b_{lk}A^{\delta b}(x)) \nonumber \\
&&+\gamma^\delta (\delta^{lk} {\overrightarrow \partial}^\mu-igT^b_{lk}A^{\nu b}(x))-\gamma^\nu (\delta^{lk} {\overleftarrow \partial}^\delta+igT^b_{lk}A^{\delta b}(x)) -\gamma^\delta (\delta^{lk} {\overleftarrow \partial}^\nu+igT^b_{lk}A^{\nu b}(x))]\psi_k(x)\nonumber \\
\label{tmnu}
\eea
where $\psi_j(x)$ is the quark field with color index $j=1,2,3$, the $A_\delta^d(x)$ is the gluon field with color index $d=1,...,8$, the Lorentz index $\delta=0,1,2,3$ and
\bea
F_{\delta \sigma}^c(x)=\partial_\delta A_\sigma^b(x) - \partial_\sigma A_\delta^b(x)+gf^{bca} A_\delta^c(x) A_\sigma^a(x).
\label{fmu}
\eea
From eq. (\ref{jmu}) we find
\bea
\frac{d}{dt}[<p,s|{\vec {\hat S}}_q|p,s>+<p,s|{\vec {\hat L}}_q|p,s>+<p,s|{\vec {\hat J}}_g|p,s>+<p,s|{\vec {\hat J}}_B|p,s>]=0
\label{ceu}
\eea
where
\bea
<p,s|{ \vec {\hat S}}_q(t) |p,s>= <p,s|\int d^3x ~{\bar \psi}_k(x)\gamma_0 {\vec \sigma} \psi_k(x)|p,s>,
\label{squ}
\eea
\bea
<p,s|{ \vec {\hat L}}_q(t)|p,s> =<p,s| \int d^3x~ {\vec x} \times {\bar \psi}_l(x) \gamma_0 [-i{\overrightarrow D}_{lk}[A]+i{\overleftarrow D}_{lk}[A]] \psi_k(x)|p,s>,
\label{lqu}
\eea
\bea
<p,s|{ \vec {\hat J}}_g(t) |p,s>= <p,s|\int d^3x~ {\vec x} \times {\vec E}^d(x) \times {\vec B}^d(x)|p,s>
\label{jgu}
\eea
and
\bea
&& <p,s|{\hat J}^n_B(t) = <p,s|\epsilon^{nsj} \int d^4x~\partial_k [x^s[E^{jc}(x)E^{kc}(x)-\delta^{kj}\frac{{E_c}^2(x)-{B_c}^2(x)}{2}]+[x^s\frac{i}{2}{\bar \psi}_{n'}(x) \nonumber \\
&&[\gamma^k(\delta_{n'n''}{\overrightarrow \partial}^j-igT^c_{n'n''}A^{jc}(x))-\gamma^k(\delta_{n'n''}{\overleftarrow \partial}^j+igT^c_{n'n''}A^{jc}(x))]\psi_{n''}(x)+\frac{1}{8}{\bar \psi}_{n'}(x)\{\gamma^n,\sigma^{sj}\}]]|p,s>.\nonumber  \\
\label{jqgu}
\eea
The covariant derivative $D_{lk}[A]$ in eq. (\ref{lqu}) is defined by
\bea
D_{ln}[A]=\delta_{ln}(i{\not \partial}-m)-igT^d_{ln}\aslash^d(x).
\eea
The ${\vec \sigma}$ in eq. (\ref{squ}) is the Pauli spin matrix and the $\sigma^{\nu \delta}$ in eq. (\ref{jqgu}) is related to the Dirac matrices via the equation
\bea
\sigma^{sl}=\frac{i}{2}[\gamma^s,\gamma^l].
\eea

From eqs. (\ref{squ}), (\ref{lqu}), (\ref{jgu}), (\ref{jqgu}) and (\ref{ceu}) we find
\bea
\frac{d}{dt}[<p,s|{{\hat S}}^z_q|p,s>+<p,s|{ {\hat L}}^z_q|p,s>+<p,s|{{\hat J}}^z_g|p,s>]=-\frac{d<p,s|{\hat J}^z_B|p,s>}{dt}
\label{xjfuxa}
\eea
where the operator ${\hat {\cal O}}^z$ means the z-component of the vector operator ${\vec {\hat {\cal O}}}$ where the expressions of the vector operators ${ \vec {\hat S}}_q(t)$, ${ \vec {\hat L}}_q(t)$, ${ \vec {\hat J}}_g(t)$ and ${ \vec {\hat J}}_B(t)$ in terms of the quark field $\psi_i(x)$ and the gluons field $A_\mu^a(x)$ are given by eqs. (\ref{squ}), (\ref{lqu}), (\ref{jgu}) and (\ref{jqgu}) respectively.

Due to the confinement of quarks and gluons inside the finite size proton we find that the quark field $\psi_i(t,r)$ and the gluon field $A_\mu^a(t,r)$ do not go to $r \rightarrow \infty$. Since the boundary surface is at the finite distance due to the finite size of the proton one finds that the boundary surface term in QCD is non-zero due to confinement of quarks and gluons inside the finite size proton irrespective of the forms of the $r$ dependence of the  quark field $\psi_i(t,r)$ and the gluon field $A_\mu^a(t,r)$ \cite{nkbsu}. Hence because of the non-zero boundary surface term in QCD due to the confinement of the quarks and gluons inside the finite size proton one finds that \cite{nkagu}
\bea
\frac{d<p,s|{\hat J}_B|p,s>}{dt}\neq 0.
\label{neq1u}
\eea
From eqs. (\ref{neq1u}) and (\ref{xjfuxa}) we find
\bea
\frac{d}{dt}[<p,s|{{\hat S}}^z_q|p,s>+<p,s|{ {\hat L}}^z_q|p,s>+<p,s|{{\hat J}}^z_g|p,s>]\neq 0
\label{xjfux}
\eea
which implies that $<p,s|{{\hat S}}^z_q|p,s>+<p,s|{ {\hat L}}^z_q|p,s>+<p,s|{{\hat J}}^z_g|p,s>$ is time dependent.

Since the proton spin $\frac{1}{2}$ is time independent and the $<p,s|{{\hat S}}^z_q|p,s>+<p,s|{ {\hat L}}^z_q|p,s>+<p,s|{{\hat J}}^z_g|p,s>$ from eq. (\ref{xjfux}) is time dependent we find that
\bea
\frac{1}{2} \neq <p,s|{{\hat S}}^z_q|p,s>+<p,s|{ {\hat L}}^z_q|p,s>+<p,s|{{\hat J}}^z_g|p,s>
\label{nequ}
\eea
which does not agree with eq. (\ref{xjfu}).

From eq. (\ref{nequ}) we find that the angular momentum sum rule in QCD as given by eq. (\ref{xjfu}) is violated due to the confinement of quarks and gluons inside the finite size proton.

\section{ Formulation of The Lattice QCD Method to study proton formation from quarks and gluons }

In this section we consider the formulation of the lattice QCD method to study the proton formation from quarks and gluons. In the next section we will extend this to study the proton spin crisis.

For the proton formation the partonic operator is given by
\bea
{\hat {\cal B}}_p(x) =\epsilon_{ijk}\psi^u_i(x) C\gamma^5 \psi^d_j(x) \psi^u_k(x)
\label{pru}
\eea
where $\psi^u_i(x)$ is the Dirac field of the up quark with color index $i=1,2,3$, the $\psi^d_i(x)$ is the Dirac field of the down quark and $C$ is the charge conjugation operator. The vacuum expectation value of the non-perturbative partonic correlation function in QCD is given by
\bea
&&<0|{\hat {\cal B}}_p(t'',r'') {\hat {\cal B}}_p(0)|0>=\frac{1}{Z[0]} \int [d{\bar \psi}^u][d{ \psi}^u] [d{\bar \psi}^d][d{ \psi}^d][dA]~{\hat {\cal B}}_p(t'',r'') {\hat {\cal B}}_p(0) \times {\rm deu}[\frac{\delta H^a_f}{\delta \omega^c}] \nonumber \\
&& \times e^{i\int d^4x [-\frac{1}{4} F_{\delta \sigma}^c(x)F^{\delta \sigma c}(x) -\frac{1}{2\alpha} [H_f^c(x)]^2 +{\bar \psi}_u^l(x)[\delta^{lk}(i{\not \partial}-m^u)+gT^c_{lk}\aslash^c(x)]\psi_u^k(x) +{\bar \psi}_d^l(x)[\delta^{lk}(i{\not \partial}-m^d)+gT^c_{lk}\aslash^c(x)]\psi_d^k(x)]} \nonumber \\
\label{cfu}
\eea
where $|0>$ is the non-perturbative QCD vacuum state ({\it i. e.}, the vacuum state of the full QCD, not of pQCD), $H_f^a(x)$ is the gauge fixing term, $\alpha$ is the gauge fixing parameter and
\bea
&&Z[0]=\int [d{\bar \psi}^u][d{ \psi}^u] [d{\bar \psi}^d][d{ \psi}^d][dA]~\times {\rm deu}[\frac{\delta H^a_f}{\delta \omega^c}] \nonumber \\
&& \times e^{i\int d^4x [-\frac{1}{4} F_{\delta \sigma}^c(x)F^{\delta \sigma c}(x) -\frac{1}{2\alpha} [H_f^c(x)]^2 +{\bar \psi}_u^l(x)[\delta^{lk}(i{\not \partial}-m^u)+gT^c_{lk}\aslash^c(x)]\psi_u^k(x) +{\bar \psi}_d^l(x)[\delta^{lk}(i{\not \partial}-m^d)+gT^c_{lk}\aslash^c(x)]\psi_d^k(x)]}\nonumber \\
 \label{zu}
\eea
is the generating functional in QCD.

The time evolution of the partonic operator in the Heisenberg representation is given by
\bea
{\hat {\cal B}}_p(t,{\vec x}) =e^{-itH}{\hat {\cal B}}_p(0,{\vec x})e^{itH}
\label{opu}
\eea
where $H$ is the QCD Hamiltonian. Inserting the complete set of energy-momentum eigenstates of the proton
\bea
\sum_{n'''} |p_{n'''}><p_{n'''}|=1
\label{csu}
\eea
in (\ref{cfu}) and then using eq. (\ref{opu}) we find in the Euclidean time that
\bea
&&\sum_{r''} e^{i{\vec p} \cdot {\vec r}''}<0|{\hat {\cal B}}_p(t'',r'') {\hat {\cal B}}_p(0)|0>=\sum_{n'''} |<0| {\hat {\cal B}}_p(0)|p_{n'''}>|^2 e^{-\int dt E_{n'''}(t)}
\label{cfux}
\eea
where $E_{n'''}(t)$ is the energy of all the quarks plus antiquarks plus gluons inside the proton and $\int dt$ is an indefinite integration. In the large Euclidean time $t\rightarrow \infty$ we find
\bea
&&\sum_{r''} e^{i{\vec p} \cdot {\vec r}''}<0|{\hat {\cal B}}_p(t'',r'') {\hat {\cal B}}_p(0)|0>|_{t''\rightarrow \infty} =|<0| {\hat {\cal B}}_p(0)|p>|^2 e^{-\int dt'' E(t'')}
\label{cfuy}
\eea

The energy of the proton $E_p$ is given by
\bea
E_p=E(t)+E_B(t)
\label{epu}
\eea
where $E_B(t)$ is non-zero boundary surface term given by \cite{nklqu}
\bea
E_B(t'') =\frac{\int d^4x'' \sum_{r'''} e^{i{\vec p} \cdot {\vec r}'''}<0|{\hat {\cal B}}_p(t''',r''') \partial_k T^{k0}(t'',{\vec x}''){\hat {\cal B}}_p(0)|0>}{\sum_{r'''} e^{i{\vec p} \cdot {\vec r}'''}<0|{\hat {\cal B}}_p(t''',r''') {\hat {\cal B}}_p(0)|0>}|_{t'''\rightarrow \infty}
\label{ebu}
\eea
where $\int dt''$ is an indefinite integration. Using eqs. (\ref{epu}) and (\ref{ebu}) in (\ref{cfuy}) we find
\bea
&&|<0| {\hat {\cal B}}_p(0)|p>|^2 e^{-tE_p}=[\frac{\sum_{r} e^{i{\vec p} \cdot {\vec r}}<0|{\hat {\cal B}}_p(t,r) {\hat {\cal B}}_p(0)|0>}{e^{\int dt\frac{\int d^4x \sum_{r'}e^{i{\vec p} \cdot {\vec r}'} <0|{\hat {\cal B}}_p(t',r') \partial_k T^{k0}(t,r){\hat {\cal B}}_p(0)|0>}{\sum_{r'} e^{i{\vec p} \cdot {\vec r}'}<0|{\hat {\cal B}}_p(t',r') {\hat {\cal B}}_p(0)|0>}|_{t'\rightarrow \infty}}}|_{t\rightarrow \infty}
\label{cfhu}
\eea
which is the equation to study the proton formation from quarks and gluons by using the lattice QCD method. This formulation of the lattice QCD method used to study various non-perturbative quantities in QCD in vacuum \cite{nklqu,nkalu} and in QCD in medium \cite{nkalu1} to study quark-gluon plasma at RHIC and LHC \cite{qgu,qgu1,qgu2,qgu3}.

\section{ Lattice QCD formulation of the Proton Spin Crisis }

From eq. (\ref{xjfuxa}) we find
\bea
\frac{d}{dt}[<p,s|{{\hat S}}^z_q|p,s>+<p,s|{ {\hat L}}^z_q|p,s>+<p,s|{{\hat J}}^z_g|p,s>+<p,s|{\hat J}^z_B|p,s>]=0
\label{xjfd}
\eea
which implies that $<p,s|{{\hat S}}^z_q|p,s>+<p,s|{ {\hat L}}^z_q|p,s>+<p,s|{{\hat J}}^z_g|p,s>+<p,s|{\hat J}^z_B|p,s>$ is time independent.

Since the spin $\frac{1}{2}$ of the proton is time independent and $<p,s|{{\hat S}}^z_q|p,s>+<p,s|{ {\hat L}}^z_q|p,s>+<p,s|{{\hat J}}^z_g|p,s>+<p,s|{\hat J}^z_B|p,s>$ from eq. (\ref{xjfd}) is time independent one finds, unlike eq. (\ref{nequ}), that
\bea
\frac{1}{2}=<p,s|{{\hat S}}^z_q|p,s>+<p,s|{ {\hat L}}^z_q|p,s>+<p,s|{{\hat J}}^z_g|p,s>+<p,s|{\hat J}^z_B|p,s>
\label{equ}
\eea
where $\frac{1}{2}$ in the left hand side is the proton spin and $<p,s|{{\hat S}}^z_q|p,s>$, $<p,s|{ {\hat L}}^z_q|p,s>$, $<p,s|{{\hat J}}^z_g|p,s>$, $<p,s|{\hat J}^z_B|p,s>$ are the z-components of $<p,s|{\vec {\hat S}}_q|p,s>$, $<p,s|{\vec {\hat L}}_q|p,s>$, $<p,s|{\vec {\hat J}}_g|p,s>$, $<p,s|{\vec {\hat J}}_B|p,s>$ where the expressions of the spin/angular momentum operators ${\vec {\hat S}}_q$, ${\vec {\hat L}}_q$, ${\vec {\hat J}}_g$, ${\vec {\hat J}}_B$ in terms of quark field $\psi_i(x)$ and gluon field $A_\mu^a(x)$ are given by eqs. (\ref{squ}), (\ref{lqu}), (\ref{jgu}) and (\ref{jqgu}) respectively.

From eq. (\ref{equ}) we find that the spin $\frac{1}{2}$ of the proton is obtained from $<p,s|{{\hat S}}^z_q|p,s>$, $<p,s|{ {\hat L}}^z_q|p,s>$, $<p,s|{{\hat J}}^z_g|p,s>$ and $<p,s|{\hat J}^z_B|p,s>$. However, these quantities $<p,s|{{\hat S}}^z_q|p,s>$, $<p,s|{ {\hat L}}^z_q|p,s>$, $<p,s|{{\hat J}}^z_g|p,s>$ and $<p,s|{\hat J}^z_B|p,s>$ are non-perturbative quantities in QCD which can not be calculated by using pQCD. On the other hand the analytical solution of the non-perturbative QCD is not known. Hence one can calculate these non-perturbative quantities $<p,s|{{\hat S}}^z_q|p,s>$, $<p,s|{ {\hat L}}^z_q|p,s>$, $<p,s|{{\hat J}}^z_g|p,s>$ and $<p,s|{\hat J}^z_B|p,s>$ by using the lattice QCD method.

In this section we will extend the lattice QCD formulation of the previous section to study the proton spin crisis, {\it i. e.}, we will formulate the lattice QCD method to calculate the non-perturbative quantities $<p,s|{{\hat S}}^z_q|p,s>$, $<p,s|{ {\hat L}}^z_q|p,s>$, $<p,s|{{\hat J}}^z_g|p,s>$ and $<p,s|{\hat J}^z_B|p,s>$ in QCD to study the proton spin crisis.

Let us first formulate the lattice QCD method to calculate the non-perturbative quantity $<p,s|{\vec {\hat S}}_q|p,s>$ in QCD where the spin operator ${\vec {\hat S}}_q$ is given by eq. (\ref{squ}) before proceeding to calculate the other non-perturbative quantities $<p,s|{\vec {\hat L}}_q|p,s>$, $<p,s|{\vec {\hat J}}_g|p,s>$ and $<p,s|{\vec {\hat J}}_B|p,s>$ in QCD where the angular momentum operators ${\vec {\hat L}}_q$, ${\vec {\hat J}}_g$ and ${\vec {\hat J}}_B$ in QCD are given by eqs. (\ref{lqu}), (\ref{jgu}) and (\ref{jqgu}) respectively.

The vacuum expectation value of the three-point non-perturbative partonic correlation function $<0|{\hat {\cal B}}_p(t'',r'') {\vec {\hat S}}_q(t'){\hat {\cal B}}_p(0)|0>$ in QCD is given by
\bea
&&<0|{\hat {\cal B}}_p(t'',r'') {\vec {\hat S}}_q(t'){\hat {\cal B}}_p(0)|0>=\frac{1}{Z[0]} \int [d{\bar \psi}^u][d{ \psi}^u] [d{\bar \psi}^d][d{ \psi}^d][dA]~{\hat {\cal B}}_p(t'',r''){\vec {\hat S}}_q(t') {\hat {\cal B}}_p(0) \times {\rm deu}[\frac{\delta H^a_f}{\delta \omega^c}] \nonumber \\
&& \times e^{i\int d^4x [-\frac{1}{4} F_{\delta \sigma}^c(x)F^{\delta \sigma c}(x) -\frac{1}{2\alpha} [H_f^c(x)]^2 +{\bar \psi}_u^l(x)[\delta^{lk}(i{\not \partial}-m^u)+gT^c_{lk}\aslash^c(x)]\psi_u^k(x) +{\bar \psi}_d^l(x)[\delta^{lk}(i{\not \partial}-m^d)+gT^c_{lk}\aslash^c(x)]\psi_d^k(x)]}
\label{3cfu}
\eea
where the partonic operator ${\vec {\hat S}}_q(t)$ is given by eq. (\ref{squ}) and the partonic operator ${\hat {\cal B}}_p(t,r) $ is given by eq. (\ref{pru}). We evaluate the ratio of the vacuum expectation value of the three-point non-perturbative partonic correlation function
\bea
C_3(p,t',t'')=\sum_{r''}e^{i{\vec p} \cdot {\vec r}''}<0|{\hat {\cal B}}_p(t'',r'') {\vec {\hat S}}_q(t'){\hat {\cal B}}_p(0)|0>
\label{3pt}
\eea
to the vacuum expectation value of the two-point non-perturbative partonic correlation function
\bea
C_2(p,t'')=\sum_{r''}e^{i{\vec p} \cdot {\vec r}''}<0|{\hat {\cal B}}_p(t'',r'') {\hat {\cal B}}_p(0)|0>
\label{2pt}
\eea
where $<0|{\hat {\cal B}}_p(t'',r'') {\vec {\hat S}}_q(t'){\hat {\cal B}}_p(0)|0>$ is given by eq. (\ref{3cfu}) and $<0|{\hat {\cal B}}_p(t'',r'') {\hat {\cal B}}_p(0)|0>$ is given by eq. (\ref{cfu}).

The complete set of energy-momentum eigenstates of the proton with spin $s$ is given by
\bea
\sum_{n'''} |p_{n'''},s><p_{n'''},s|=1.
\label{csus}
\eea
Using eqs. (\ref{opu}) and (\ref{csus}) in
\bea
\frac{C_3(p,t',t'')}{C_2(p,t'')}=\frac{\sum_{r''}e^{i{\vec p} \cdot {\vec r}''}<0|{\hat {\cal B}}_p(t'',r'') {\vec {\hat S}}_q(t'){\hat {\cal B}}_p(0)|0>}{\sum_{r''}e^{i{\vec p} \cdot {\vec r}''}<0|{\hat {\cal B}}_p(t'',r'') {\hat {\cal B}}_p(0)|0>}
\label{rpt}
\eea
we find in Euclidean time
\bea
\frac{C_3(p,t',t'')}{C_2(p,t'')}=\frac{\sum_{n'',n'''}<0|{\hat {\cal B}}_p(0)|p_{n'''},s><p_{n'''},s| {\vec {\hat S}}_q(t') |p_{n''},s><p_{n''},s|{\hat {\cal B}}_p(0)|0>e^{-\int dt''' E_{n'''}(t''')}}{\sum_{n'''}|<0|{\hat {\cal B}}_p(0)|p_{n'''},s>|^2 e^{-\int dt''' E_{n'''}(t''')}}.\nonumber \\
\label{rptx}
\eea
Neglecting the higher energy level contributions at the large time limit $t'' \rightarrow \infty$ we find
\bea
\frac{C_3(p,t',t'')}{C_2(p,t'')}|_{t''\rightarrow \infty} =\frac{|<0|{\hat {\cal B}}_p(0)|p,s>|^2<p,s| {\vec {\hat S}}_q(t') |p,s>e^{-\int dt''' E(t''')}}{|<0|{\hat {\cal B}}_p(0)|p,s>|^2 e^{-\int dt''' E(t''')}}
\label{rpty}
\eea
where for the ground state $n'''=n''=0$ of the proton we have
\bea
|p_0,s>=|p,s>,~~~~~~~~~E_0(t)=E(t).
\eea
From eq. (\ref{rpty}) and (\ref{rpt}) we find
\bea
<p,s| {\vec {\hat S}}_q(t) |p,s>=\frac{\sum_{r'}e^{i{\vec p} \cdot {\vec r}'}<0|{\hat {\cal B}}_p(t',r') {\vec {\hat S}}_q(t){\hat {\cal B}}_p(0)|0>}{\sum_{r'}e^{i{\vec p} \cdot {\vec r}'}<0|{\hat {\cal B}}_p(t',r') {\hat {\cal B}}_p(0)|0>}|_{t'\rightarrow \infty}
\label{rptz}
\eea
where the partonic operator ${\vec {\hat S}}_q(t)$ is given by eq. (\ref{squ}), the partonic operator ${\hat {\cal B}}_p(t,r) $ is given by eq. (\ref{pru}), the vacuum expectation value of the two-point non-perturbative partonic correlation function $<0|{\hat {\cal B}}_p(t',r') {\hat {\cal B}}_p(0)|0>$ is given by eq. (\ref{cfu}) and the vacuum expectation value of the three-point non-perturbative partonic correlation function $<0|{\hat {\cal B}}_p(t',r') {\vec {\hat S}}_q(t){\hat {\cal B}}_p(0)|0>$ is given by eq. (\ref{3cfu}).

Since the vacuum expectation value of the two-point non-perturbative partonic correlation function $<0|{\hat {\cal B}}_p(t',r') {\hat {\cal B}}_p(0)|0>$ in eq. (\ref{cfu}) and the vacuum expectation value of the three-point non-perturbative partonic correlation function $<0|{\hat {\cal B}}_p(t',r') {\vec {\hat S}}_q(t){\hat {\cal B}}_p(0)|0>$ in eq. (\ref{3cfu}) can be calculated by using the lattice QCD method we find that the non-perturbative quantity $<p,s| {\vec {\hat S}}_q(t) |p,s>$ in eq. (\ref{equ}) to study the proton spin crisis can be calculated from eq. (\ref{rptz}) by using the lattice QCD method.

Following the similar procedure we find
\bea
<p,s| {\vec {\hat L}}_q(t) |p,s>=\frac{\sum_{r'}e^{i{\vec p} \cdot {\vec r}'}<0|{\hat {\cal B}}_p(t',r') {\vec {\hat L}}_q(t){\hat {\cal B}}_p(0)|0>}{\sum_{r'}e^{i{\vec p} \cdot {\vec r}'}<0|{\hat {\cal B}}_p(t',r') {\hat {\cal B}}_p(0)|0>}|_{t'\rightarrow \infty}
\label{rptz1}
\eea
where the partonic operator ${\vec {\hat L}}_q(t)$ is given by eq. (\ref{lqu}), the partonic operator ${\hat {\cal B}}_p(t,r) $ is given by eq. (\ref{pru}), the vacuum expectation value of the two-point non-perturbative partonic correlation function $<0|{\hat {\cal B}}_p(t',r') {\hat {\cal B}}_p(0)|0>$ is given by eq. (\ref{cfu}) and the vacuum expectation value of the three-point non-perturbative partonic correlation function $<0|{\hat {\cal B}}_p(t',r') {\vec {\hat L}}_q(t){\hat {\cal B}}_p(0)|0>$ is given by
\bea
&&<0|{\hat {\cal B}}_p(t'',r'') {\vec {\hat L}}_q(t'){\hat {\cal B}}_p(0)|0>=\frac{1}{Z[0]} \int [d{\bar \psi}^u][d{ \psi}^u] [d{\bar \psi}^d][d{ \psi}^d][dA]~{\hat {\cal B}}_p(t'',r''){\vec {\hat L}}_q(t') {\hat {\cal B}}_p(0) \times {\rm deu}[\frac{\delta H^a_f}{\delta \omega^c}] \nonumber \\
&& \times e^{i\int d^4x [-\frac{1}{4} F_{\delta \sigma}^c(x)F^{\delta \sigma c}(x) -\frac{1}{2\alpha} [H_f^c(x)]^2 +{\bar \psi}_u^l(x)[\delta^{lk}(i{\not \partial}-m^u)+gT^c_{lk}\aslash^c(x)]\psi_u^k(x) +{\bar \psi}_d^l(x)[\delta^{lk}(i{\not \partial}-m^d)+gT^c_{lk}\aslash^c(x)]\psi_d^k(x)]}.
\label{l3cfu}
\eea
Since the vacuum expectation value of the two-point non-perturbative partonic correlation function $<0|{\hat {\cal B}}_p(t',r') {\hat {\cal B}}_p(0)|0>$ in eq. (\ref{cfu}) and the vacuum expectation value of the three-point non-perturbative partonic correlation function $<0|{\hat {\cal B}}_p(t',r') {\vec {\hat L}}_q(t){\hat {\cal B}}_p(0)|0>$ in eq. (\ref{l3cfu}) can be calculated by using the lattice QCD method we find that the non-perturbative quantity $<p,s| {\vec {\hat L}}_q(t) |p,s>$ in eq. (\ref{equ}) to study the proton spin crisis can be calculated from eq. (\ref{rptz1}) by using the lattice QCD method.

Similarly we find
\bea
<p,s| {\vec {\hat J}}_g(t) |p,s>=\frac{\sum_{r'}e^{i{\vec p} \cdot {\vec r}'}<0|{\hat {\cal B}}_p(t',r') {\vec {\hat J}}_g(t){\hat {\cal B}}_p(0)|0>}{\sum_{r'}e^{i{\vec p} \cdot {\vec r}'}<0|{\hat {\cal B}}_p(t',r') {\hat {\cal B}}_p(0)|0>}|_{t'\rightarrow \infty}
\label{rptz2}
\eea
where the partonic operator ${\vec {\hat J}}_g(t)$ is given by eq. (\ref{jgu}), the partonic operator ${\hat {\cal B}}_p(t,r) $ is given by eq. (\ref{pru}), the vacuum expectation value of the two-point non-perturbative partonic correlation function $<0|{\hat {\cal B}}_p(t',r') {\hat {\cal B}}_p(0)|0>$ is given by eq. (\ref{cfu}) and the vacuum expectation value of the three-point non-perturbative partonic correlation function $<0|{\hat {\cal B}}_p(t',r') {\vec {\hat J}}_g(t){\hat {\cal B}}_p(0)|0>$ is given by
\bea
&&<0|{\hat {\cal B}}_p(t'',r'') {\vec {\hat J}}_g(t'){\hat {\cal B}}_p(0)|0>=\frac{1}{Z[0]} \int [d{\bar \psi}^u][d{ \psi}^u] [d{\bar \psi}^d][d{ \psi}^d][dA]~{\hat {\cal B}}_p(t'',r''){\vec {\hat J}}_g(t') {\hat {\cal B}}_p(0) \times {\rm deu}[\frac{\delta H^a_f}{\delta \omega^c}] \nonumber \\
&& \times e^{i\int d^4x [-\frac{1}{4} F_{\delta \sigma}^c(x)F^{\delta \sigma c}(x) -\frac{1}{2\alpha} [H_f^c(x)]^2 +{\bar \psi}_u^l(x)[\delta^{lk}(i{\not \partial}-m^u)+gT^c_{lk}\aslash^c(x)]\psi_u^k(x) +{\bar \psi}_d^l(x)[\delta^{lk}(i{\not \partial}-m^d)+gT^c_{lk}\aslash^c(x)]\psi_d^k(x)]}.
\label{j3cfu}
\eea
Since the vacuum expectation value of the two-point non-perturbative partonic correlation function $<0|{\hat {\cal B}}_p(t',r') {\hat {\cal B}}_p(0)|0>$ in eq. (\ref{cfu}) and the vacuum expectation value of the three-point non-perturbative partonic correlation function $<0|{\hat {\cal B}}_p(t',r') {\vec {\hat J}}_g(t){\hat {\cal B}}_p(0)|0>$ in eq. (\ref{j3cfu}) can be calculated by using the lattice QCD method we find that the non-perturbative quantity $<p,s| {\vec {\hat J}}_g(t) |p,s>$ in eq. (\ref{equ}) to study the proton spin crisis can be calculated from eq. (\ref{rptz2}) by using the lattice QCD method.

Finally we find \cite{nkncp}
\bea
<p,s| {\vec {\hat J}}_B(t) |p,s>=\frac{\sum_{r'}e^{i{\vec p} \cdot {\vec r}'}<0|{\hat {\cal B}}_p(t',r') {\vec {\hat J}}_B(t){\hat {\cal B}}_p(0)|0>}{\sum_{r'}e^{i{\vec p} \cdot {\vec r}'}<0|{\hat {\cal B}}_p(t',r') {\hat {\cal B}}_p(0)|0>}|_{t'\rightarrow \infty}
\label{rptz3}
\eea
where the partonic operator ${\vec {\hat J}}_B(t)$ is given by eq. (\ref{jqgu}), the partonic operator ${\hat {\cal B}}_p(t,r) $ is given by eq. (\ref{pru}), the vacuum expectation value of the two-point non-perturbative partonic correlation function $<0|{\hat {\cal B}}_p(t',r') {\hat {\cal B}}_p(0)|0>$ is given by eq. (\ref{cfu}) and the vacuum expectation value of the three-point non-perturbative partonic correlation function $<0|{\hat {\cal B}}_p(t',r') {\vec {\hat J}}_B(t){\hat {\cal B}}_p(0)|0>$ is given by
\bea
&&<0|{\hat {\cal B}}_p(t'',r'') {\vec {\hat J}}_B(t'){\hat {\cal B}}_p(0)|0>=\frac{1}{Z[0]} \int [d{\bar \psi}^u][d{ \psi}^u] [d{\bar \psi}^d][d{ \psi}^d][dA]~{\hat {\cal B}}_p(t'',r''){\vec {\hat J}}_B(t') {\hat {\cal B}}_p(0) \times {\rm deu}[\frac{\delta H^a_f}{\delta \omega^c}] \nonumber \\
&& \times e^{i\int d^4x [-\frac{1}{4} F_{\delta \sigma}^c(x)F^{\delta \sigma c}(x) -\frac{1}{2\alpha} [H_f^c(x)]^2 +{\bar \psi}_u^l(x)[\delta^{lk}(i{\not \partial}-m^u)+gT^c_{lk}\aslash^c(x)]\psi_u^k(x) +{\bar \psi}_d^l(x)[\delta^{lk}(i{\not \partial}-m^d)+gT^c_{lk}\aslash^c(x)]\psi_d^k(x)]}.
\label{jqg3cfu}
\eea
Since the vacuum expectation value of the two-point non-perturbative partonic correlation function $<0|{\hat {\cal B}}_p(t',r') {\hat {\cal B}}_p(0)|0>$ in eq. (\ref{cfu}) and the vacuum expectation value of the three-point non-perturbative partonic correlation function $<0|{\hat {\cal B}}_p(t',r') {\vec {\hat J}}_B(t){\hat {\cal B}}_p(0)|0>$ in eq. (\ref{jqg3cfu}) can be calculated by using the lattice QCD method we find that the non-perturbative quantity $<p,s| {\vec {\hat J}}_B(t) |p,s>$ in eq. (\ref{equ}) to study the proton spin crisis can be calculated from eq. (\ref{rptz3}) by using the lattice QCD method.

Hence we find that the spin $\frac{1}{2}$ of the proton in eq. (\ref{equ}) can be calculated by using the lattice QCD method using the non-perturbative formulas derived in eqs. (\ref{rptz}), (\ref{rptz1}), (\ref{rptz2}), (\ref{rptz3}) where the spin/angular momentum operators of the partons in QCD are given by eqs. (\ref{squ}), (\ref{lqu}), (\ref{jgu}), (\ref{jqgu}) respectively with the partonic operator ${\hat {\cal B}}_p(x)$ given by $(\ref{pru})$.

\section{Conclusions}
The proton spin crisis remains an unsolved problem in particle physics. The spin and angular momentum of the partons inside the proton are non-perturbative quantities in QCD which cannot be calculated by using the perturbative QCD (pQCD). In this paper we have presented the lattice QCD formulation to study the proton spin crisis. We have derived the non-perturbative formula of the spin and angular momentum of the partons inside the proton from the first principle in QCD which can be calculated by using the lattice QCD method.


\begin{thebibliography}{99}

\bibitem{emu} J. Ashman {\it et al}, European Muon Collaboration, Phys. Lett. B 206 (1988) 364; Nucl. Phys. B 328 (1989) 1.

\bibitem{ohu} {\it See for example}, C. A. Aidala, S. D. Bass, D. Hasch and G. K. Mallot, Rev. Mod. Phys. 85 (2013) 655.

\bibitem{hcu} A. Adare {\it et al.}, PHENIX Collaboration, Phys. Rev. D 93 (2016) 011501; A. Adare {\it et al.}, PHENIX Collaboration, Phys. Rev. D 94 (2016) 112008.

\bibitem{wdu} A. Bazilevsky, J. Phys. Conf. Ser. 678 (2016) 012059.

\bibitem{jfu} J. R. Ellis and R. L. Jaffe, Phys. Rev. D 9 (1974) 1444; Phys. Rev. D 10 (1974) 1669; R. L. Jaffe and A. Manohar, Nucl. Phys. B337 (1990) 509; A. V. Manohar, Phys. Rev. Lett. 66 (1991) 289.

\bibitem{xfu} X.-D. Ji, Phys. Rev. Lett. 78 (1997) 610.

\bibitem{nkymu} G. C. Nayak, arXiv:1802.07825.

\bibitem{nkbsu} G. C. Nayak, arXiv:1807.09158.

\bibitem{nkagu} G. C. Nayak, arXiv:1803.08371 [hep-ph].

\bibitem{nklqu} G. C. Nayak, arXiv:1811.09685.

\bibitem{nkalu} G. C. Nayak, arXiv:1810.12088;  arXiv:1904.03998.

\bibitem{nkalu1} G. C. Nayak, . arXiv:1902.10522; arXiv:1904.05376.


\bibitem{qgu} G. C. Nayak and P. van Nieuwenhuizen, Phys. Rev. D 71 (2005) 125001; G. C. Nayak, Phys. Rev. D 72 (2005) 125010; F. Cooper and G. C. Nayak, Phys. Rev. D73 (2006) 065005; D. Dietrich, G. C. Nayak and W. Greiner, Phys. Rev. D64 (2001) 074006; M. C. Birse, C-W. Kao and G. C. Nayak, Phys. Lett. B570 (2003) 171.

\bibitem{qgu1} G. C. Nayak {\it et al.}, Nucl. Phys. A687 (2001) 457; G. C. Nayak and R. S. Bhalerao, Phys. Rev. C 61 (2000) 054907; G. C. Nayak and V. Ravishankar, Phys. Rev. C 58 (1998) 356; Phys. Rev. D 55 (1997) 6877; F. Cooper, E. Mottola and G. C. Nayak, Phys. Lett. B555 (2003) 181.

\bibitem{qgu2} G. C. Nayak, Annals Phys. 325 (2010) 682; Phys. Lett. B442 (1998) 427; JHEP 9802 (1998) 005; Eur. Phys. J. C64 (2009) 73; JHEP 0906 (2009) 071; Annals Phys. 324 (2009) 2579; Annals Phys. 325 (2010) 514; Eur. Phys. J.C59 (2009) 715.

\bibitem{qgu3} G. C. Nayak, Eur. Phys. J. Plus 133 (2018) 52; arXiv:1506.02651 [hep-ph]; J. Theor. Appl. Phys. 11 (2017) 275; arXiv:1705.07913 [hep-ph]; G. C. Nayak, JHEP 1709 (2017) 090; arXiv:1808.01937; Eur. Phys. J. C76 (2016) 448.

\bibitem{nkncp} G. C. Nayak, arXiv:1905.03717 [hep-ph].


\end{thebibliography}
\end{document}